\documentclass[article]{elsarticle}
\usepackage{verbatim}%FOR "COMMENT" ENVIRONMENT
\usepackage{amsfonts}
\usepackage[centertags]{amsmath}
\usepackage{amssymb}
\usepackage{amsthm}
\usepackage{wasysym}
\usepackage{amsbsy}
\usepackage{graphics}
\usepackage{textcomp}
\usepackage{color} 
\usepackage{lineno,hyperref} 
\usepackage{setspace} 
\usepackage{wrapfig} 
\usepackage{sidecap} 
\usepackage{eurosym}
\usepackage{enumitem} 
\usepackage{multirow}  
\usepackage[dvipsnames,table]{xcolor} 
\usepackage{array}
\modulolinenumbers[1]

\journal{NIMA}
\usepackage{amsmath,amsfonts,amssymb}
\usepackage{graphicx} 
\usepackage{doi}
\begin{document}

\begin{frontmatter}

\title{Monte-Carlo simulation of neutron transmission through nanocomposite materials for neutron-optics applications}

\author[AIT]{M. Blaickner} 
\author[Atominstitut]{B. Demirel}
\author[IJS,uniLjubljana]{I. Dreven\v{s}ek-Olenik} 
\author[uniwien]{M. Fally} 
\author[uniwien]{P. Flauger}
\author[ILL]{P. Geltenbort} 
\author[Atominstitut]{Y. Hasegawa}   
\author[Atominstitut]{R. Kurinjimala} 
\author[IJS]{M. Li\v{c}en} 
\author[uniSalzburg]{C. Pruner}  
\author[Atominstitut]{S. Sponar} 
\author[uniTokyo]{Y. Tomita}  
\author[uniwien]{J. Klepp\corref{mycorrespondingauthor}}
\ead{juergen.klepp@univie.ac.at}
\cortext[mycorrespondingauthor]{Corresponding author}

\address[AIT]{AIT Austrian Institute of Technology, 2444 Seibersdorf, Austria} 
\address[Atominstitut]{Atominstitut, TU Vienna, Stadionallee 2, 1020 Vienna, Austria} 
\address[IJS]{Jo\v{z}ef Stefan Institute, Department of Complex Matter, Jamova 39, SI-1000 Ljubljana, Slovenia} 
\address[uniLjubljana]{University of Ljubljana, Faculty of Mathematics and Physics, Jadranska 19, SI-1000 Ljubljana, Slovenia}
\address[uniwien]{Faculty of Physics, Boltzmanngasse 5, University of Vienna, 1090 Vienna, Austria}
\address[ILL]{Institut Laue-Langevin, 71 avenue des Martyrs, 38000 Grenoble, France} 
\address[uniSalzburg]{Department of Chemistry and Physics of Materials, Jakob-Haringer-Strasse 2 a, University of Salzburg, 5020 Salzburg, Austria} 
\address[uniTokyo]{Department of Engineering Science, University of Electro-Communications, Chofu, Tokyo 182-8585, Japan}

\begin{abstract}
Nanocomposites enable us to tune parameters that are crucial for use of such materials for neutron-optics applications. By careful choice of properties such as species (isotope) and concentration of contained nanoparticles, diffractive optical elements for long-wavelength neutrons are feasible. 
Nanocomposites for neutron optics have so far been tested successfully in protonated form, containing high amounts of $^1$H atoms, which exhibit rather strong neutron absorption and incoherent scattering. At a future stage of development,  chemicals containing $^1$H could be replaced by components containing more favourable isotopes, such as $^2$H or $^{19}$F. 
In this note, we present results of Monte-Carlo simulations of the transmissivity of various nanocomposite materials for thermal and very-cold neutron spectra. Our simulation results for deuterated and fluorinated nanocomposite materials predict the losses due to absorption and scattering to be as low as 2\%, as well as the broadening of the beam cross section to be negligible.
%
%
%a decrease of absorption- and scattering-losses down to about 2\,\% for very-cold neutrons and negligible broadening of the beam cross section. 
\end{abstract}


\end{frontmatter}

%\linenumbers

\section{Introduction}\label{sec:intro} 

\begin{figure}
   \begin{center}
      \includegraphics[height=5.7cm]{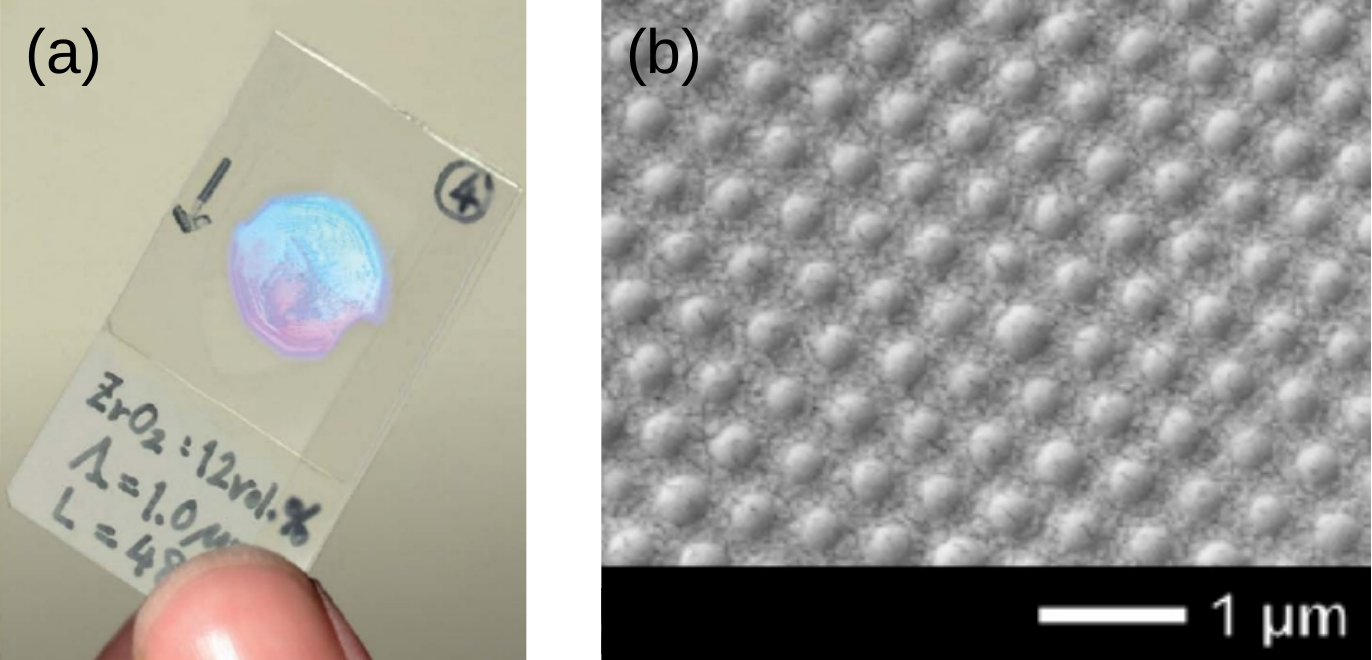}
   \end{center}
\caption[example] 
{(a) Light diffraction from a holographic ZiO$_2$-nanoparticle-polymer composite grating under white light illumination. The arrow denotes the direction of the grating vector. (b) SEM image of a colloidal crystal containing polystyrene beads and superparamagnetic maghemite nanoparticles. \label{fig1}}
\end{figure} 
Experiments with holographically produced neutron-optical gratings containing deuterated (poly)methylmethacrylate lead to successful tests of triple-Laue interferometers for cold neutrons \cite{schellhornPhysB1997,prunerNIMA2006}. More recent data suggest that nanocomposites can be tailored for particular applications in neutron optics due to their flexibility regarding choice of material and grating parameters.
 
In a nutshell, in the case of photopolymerizable nanoparticle-polymer composites, mixtures of monomer, photoinitiator chemicals and surface-treated nanoparticles (average core diameter in the range of 10\,nm \cite{suzukiAO2004}) are exposed to a light intensity pattern with a periodicity of several 100\,nm. The polymerization in bright regions leads to mutual diffusion and phase separation of monomer and nanoparticles under holographic exposure, forming gratings of 50 - 100\,$\mu$m thickness [see Fig.\,\ref{fig1}(a)]. The decisive neutron-optical features of such gratings can be chosen so that the gratings work as beam splitters, three-port beam splitters or mirrors for long-wavelength neutrons as demonstrated in Refs.\,\cite{fallyPRL2010,kleppMaterials2012,kleppPTEP2014,tomitaJOMO2016}. 
Furthermore, colloidal crystals loaded with superparamagnetic nanoparticles [see Fig.\,\ref{fig1}(b)] have recently been demonstrated to exhibit varying neutron-polarization properties as a function of the external magnetic field \cite{licenJPCS2017}, which might pave the way to inexpensive, business-card size polarizing beam splitters for neutrons. Recently, we expanded our search for suitable materials further to include mixtures containing ionic liquids \cite{ellabbanMaterials2017}. 

The above mentioned nanocomposite materials have in common that a large part of their volume is made up of chemicals which contain a considerable percentage of hydrogen atoms ($^1$H). Owing to the low thickness of neutron-optical nanocomposite components, all experiments carried out so far have been successful, as around 25\,\% transmission loss (for very-cold neutrons, see below) is tolerable in proof-of-principle experiments. However, the idea to replace chemical constituents containing $^1$H by ingredients containing $^2$H or $^{19}$F suggests itself and is certainly to be considered once development of neutron-optical components for day-to-day operation is aimed at. In the present paper, we estimate the potential impact of this approach.

Monte Carlo simulations have been proven to be very useful in describing neutron transport. In particular the Monte Carlo N-Particle (MCNP) transport code developed at the Los Alamos National Laboratory \cite{goorleyNT2012} has served multiple purposes in neutron physics, from simulating ultra-cold neutron sources \cite{karchEPJA2014} to applications in Neutron Capture Therapy \cite{blaicknerARI2012, ziegnerMP2014, schmitzMP2015, petersRO2015}. In this study MCNP6 was used to calculate neutron transmissivity with regard to different isotope compositions of nanocomposite materials for neutron optics. 

\section{Materials and methods}\label{sec:Meth} 
 
\subsection{Simulations}\label{subsec:Simu} 

An MCNP model was implemented, representing a slab of nanocomposite material of 100\,$\mu$m thickness at room temperature through which neutrons of certain wavelengths are travelling. 20\,\% of its volume consists of dispersed SiO$_2$ nanoparticles with the core density of 2.6\,g/cm$^3$ whereas the matrix consists of polymethylmethacrylate (PMMA) or polydimethylsiloxane (PDMS) with densities of about 1.2\,g/cm$^3$ and 0.965\,g/cm$^3$, respectively. Four different versions of matrix materials were incorporated into the model, (\emph{i}) the conventional PMMA (C$_5$O$_2$H$_8$)$_n$, (\emph{ii}) its deuterated form (C$_5$O$_2$D$_8$)$_n$, (\emph{iii}) a fluorinated form (C$_5$O$_2$F$_8$)$_n$, (\emph{iv}) PDMS (C$_2$H$_6$OSi)$_n$ (polymer-chain ends Si(CH$_3$)$_3$ were neglected due to the typically very large number of repeating monomer units $n$), (\emph{v}) deuterated PDMS (C$_2$D$_6$OSi)$_n$ and (\emph{vi}) a fluorinated PDMS (C$_2$F$_6$OSi)$_n$. Here, D is referring to the isotope $^2$H. Note that the three different versions of ``PMMA'' and ``PDMS'' are used for the sake of simplicity. Similar to current nanocomposite materials used in neutron optics, the actual realization of fluorinated nanocomposites will somewhat differ from the chemical formulas stated above. Nevertheless, the isotopic composition of future materials may be expected to be similar to the ones used in the model substances.

The elemental compositions of the six model-substances were treated as homogeneous mixtures of the stable isotopes' natural abundances. The materials are surrounded by air in its natural, elemental composition. Diffraction is not taken into account in our simulations as it also does not play a role in the experiments (see below). 

Energy distributions of the incoming neutrons were computed based on the fits to the measured time-of-flight data (TOF, see below) in Fig.\,\ref{fig2} after subtraction of experimental background.
The neutron source was modeled as a surface of  $1 \times 1$\,mm$^2$ and is placed 1\,cm in front of the grating. The source plane is parallel to the sample's front and back surfaces, with the optical axis going through the center of both surfaces. The incident neutron spectrum was implemented as being emitted mono-directionally, i.e. orthogonal to the source plane and in direction of the material slab---a hypothetical, perfectly collimated beam. The neutron flux was then simulated at two planes, both of which are parallel to the sample's front and back surfaces and are situated 100\,$\mu$m in front of the sample and 100\,$\mu$m behind the sample. Additionally, the percentage of transmitted neutrons proceeding into the solid angle of a narrow circular cone in forward direction was calculated. In order to achieve good statistics 50 million neutrons were simulated in each run, using the ENDF60 library for neutron cross section data \cite{mcfarlandLANLRep1994}.    
Due to the number of simulated particles the relative error of the neutron fluxes calculated by MCNP tends to zero in all cases, indicating ideal statistics for the given simulation scenario. 
%The results of the simulations are given in Table\,\ref{tab:Results}.

\subsection{Experiments}\label{subsec:Exp} 
\begin{figure}
   \begin{center}
      \includegraphics[height=4.4cm]{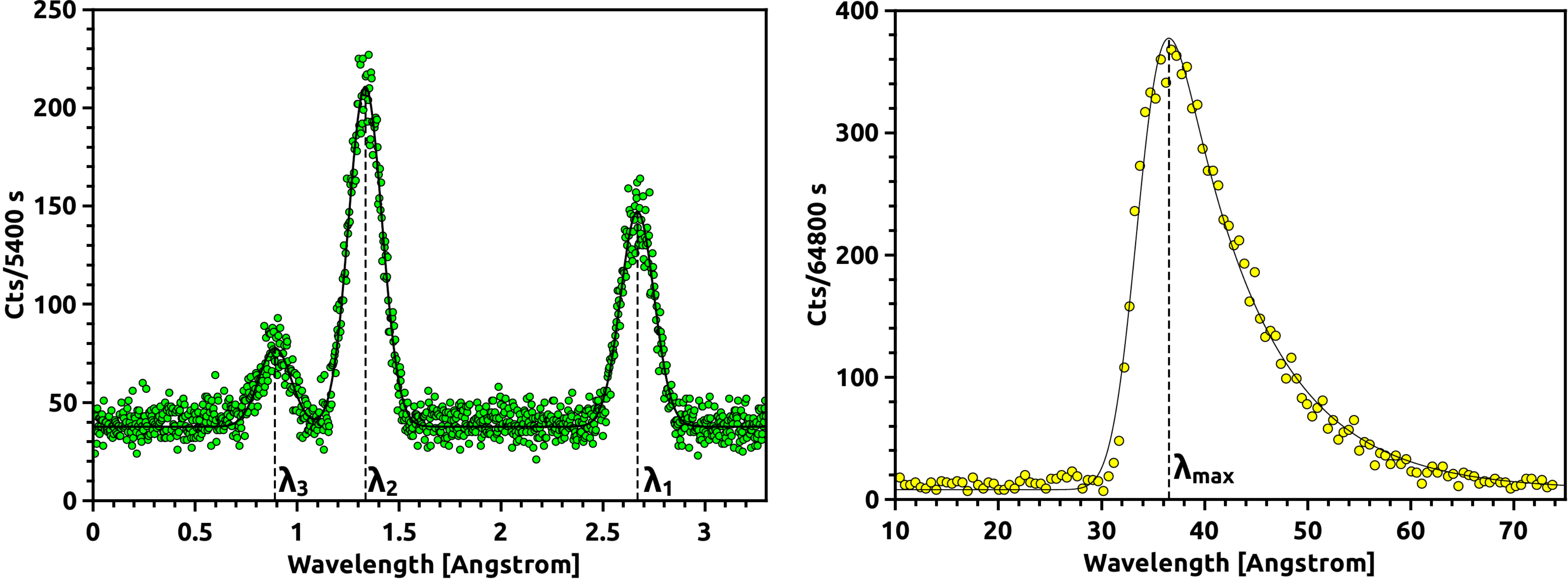}
   \end{center}
\caption[example] 
{(a) TOF data (thermal neutrons) of the spectrum used for transmission measurements at the TRIGA reactor of the Atominstitut, Vienna University of Technology, Austria. The solid line is a multi-peak Gaussian fit to the data used to compute the energy spectrum for MCNP simulations. (b) TOF data (very-cold neutrons) of the spectrum used for transmission measurements at the Institut Laue Langevin (ILL), Grenoble, France. The solid line is an exponentially modified Gaussian fit to the data used to calculate the energy spectrum for MCNP simulations.\label{fig2}}
\end{figure}

\paragraph{Thermal neutrons} 
The spectrum of the neutron beam we used for transmission measurements at the TRIGA reactor of the Atominstitut of Vienna University of Technology was recorded by the TOF method. In TOF measurements, a beam chopper cuts a continuous beam into neutron pulses. Synchronous to the chopping, a voltage signal triggers electronic sorting of detected neutrons of each pulse according to their arrival times. A thin pencil-shaped $^3$He-gas detector ($\approx 2\,$cm in diameter) was used for reasonable time resolution. The raw data and a multi-peak Gaussian fit is shown in Fig.\,\ref{fig2}\,(a). As can be seen, background noise was considerable. However, this does not hinder the determination of three peak positions and their relative intensities at a satisfactory level of precision. An additional detector was mounted near the beam port exit to correct for small reactor power fluctuations. The appearance of three peaks is due to the setting of the pyrolithic graphite monochromator crystal of the beamline and the reactor spectrum: The Bragg equation $n\lambda_n = 2 d \sin\theta_B$ (with an integer $n$, the neutron wavelength $\lambda$, the crystal plane constant $d$ and Bragg-angle $\theta_B$) is fulfilled for several wavelengths emitted by the reactor at given $\theta_B$ and $n=1,2,3$. From the fit-parameter estimations for the largest peak and the Bragg condition, we found these wavelengths to be $\lambda_1=2.667(64)$\,\AA, $\lambda_2=1.334(32)$\,\AA, 
$\lambda_3=0.889(22)$\,\AA, corresponding to neutron energies $E_1\approx 12$\,meV, $E_2\approx 46$\,meV and $E_3\approx 104$\,meV, respectively. 

The sample was a holographic nanoparticle-polymer grating \cite{tomitaJOMO2016} of about 100\,$\mu$m thickness including 20\,vol\% of SiO$_2$ nanoparticles. It contained hydrogen isotopes in natural elemental composition (a protonated sample). Due to flux limitations of the neutron source, the transmissivity of the sample grating was measured without chopper, with the detector at about 10\,cm distance behind the sample. The measured values include the transmissivity of two 1-mm thick covering glass plates, one at the front- and the other at the back-surface of the grating. The sample was mounted on a Cd-aperture of dimensions $5.1\times 6.5$\,mm$^2$. Using additional transmissivity measurements of the aperture without sample as well as of a single glass plate and a measurement of the dark counts, we calculated the net transmissivity $T_{\text{exp}}$ of the nanocomposite material.

\paragraph{Very-cold neutrons (VCN)}
The spectrum of the VCN beam of the instrument PF2 of the ILL, Grenoble, France was also measured using the TOF method. The data are shown in Fig.\,\ref{fig2}\,(b). In this case, a $^3$He area-detector with a pixel size of about $2\times$2\,mm$^2$ was available. We selected the beam part of higher flux, but smaller wavelength of the VCN beam of PF2 \cite{odaNIMA2017}. The peak wavelength $\lambda_{\text{\scriptsize{max}}}$ was about 37\,\AA\,, which was also checked by diffraction from a nanocomposite grating of known grating period. One can calculate the relative width of the wavelength distribution to be as large as $\Delta\lambda/\lambda\approx 0.4$, where $\Delta\lambda$ is approximately the FWHM of the asymmetric distribution in Fig.\,\ref{fig2}\,(b). The beam geometry was set for diffraction experiments, using small Cd-apertures of $1\times 8\,$mm$^2$ resulting in a horizontally well collimated beam (roughly 0.001\,mrad divergence) at large sample-detector distance of about 2\,m. Sealed plastic tubes, constantly flushed with helium gas, were used to bridge long flight paths in air to avoid detrimental scattering of neutrons in air. Transmission measurements were done with a holographic nanoparticle-polymer grating \cite{tomitaJOMO2016} with 25\,vol\% SiO$_2$ nanoparticles and 100\,$\mu$m thickness, containing H isotopes in natural elemental composition, the nanoparticle concentration being only slightly different from the one used in the experiment with thermal neutrons. The difference does not result in substantial differences in $T_\text{exp}$ values, as can be seen from the results (see below and Table\,\ref{tab:Results}). 

\vspace*{3mm} 

\noindent Note that the samples for the transmission measurements embodied a grating structure. Diffraction from this grating structure can be neglected because we choose the angle of incidence far from the Bragg-position or integrate over all diffraction spots that might appear. Therefore, diffraction from the grating structure is not accounted for in the simulation model.

\section{Results and discussion}\label{sec:ResAndDisc} 

%\begin{table}[] 
%\begin{center}
%\begin{tabular}{|m{2.cm}| l l l l|}
%\hline
%\multicolumn{ 1}{|c|}{} & \multicolumn{ 1}{c|}{$T_{\text{exp}}$ (thermal)} & \multicolumn{ 1}{c|}{$T_{\text{sim}}$ (thermal)} & \multicolumn{ 1}{c|}{$T_{\text{exp}}$ (VCN)} & \multicolumn{ 1}{c|}{$T_{\text{sim}}$ (VCN)} \\ \hline 
%\multicolumn{ 1}{|c|}{$^1$H} & \multicolumn{ 1}{c|}{$\approx 0.99$} & \multicolumn{ 1}{c|}{0.99} & \multicolumn{ 1}{c|}{$\approx 0.73$} & \multicolumn{ 1}{c|}{0.77} \\ \hline 
%\multicolumn{ 1}{|c|}{D} & \multicolumn{ 1}{c|}{-} & \multicolumn{ 1}{c|}{-} & \multicolumn{ 1}{c|}{$\approx 0.96^\dagger$} & \multicolumn{ 1}{c|}{0.95} \\ \hline
%\multicolumn{ 1}{|c|}{$^{19}$F} & \multicolumn{ 1}{c|}{-} & \multicolumn{ 1}{c|}{-} & \multicolumn{ 1}{c|}{-} & \multicolumn{ 1}{c|}{0.98} \\ \hline
%\end{tabular} 
%\caption{Results of experiments ($T_{\text{exp}}$) and simulations ($T_{\text{sim}}$) for nanocomposite model materials of 100\,$\mu$m thickness containing $^1$H, D or $^{19}$F for the thermal and very cold spectra of Fig.\,\ref{fig2}. Numbers with preceding `$\approx$' are measured values. Relative measurement uncertainties are estimated to be up to 5\,\%. Empty cells correspond to values close to unity, that were neither measured (due to lack of sample material) nor computed. ($^\dagger$This value was derived from a measured value in Ref.\,\cite{prunerNIMA2006}. See text for explanations.)}
%\label{tab:Results}
%\end{center} 
%\end{table}

\begin{table}[] 
\begin{center}
\begin{tabular}{|m{2.cm}|l l l|}
\hline
\multicolumn{ 1}{|c|}{Isotope} & \multicolumn{ 1}{c|}{$T_{\text{exp}}$ (PMMA)} & \multicolumn{ 1}{c|}{$T_{\text{sim}}$ (PMMA)} & \multicolumn{ 1}{c|}{$T_{\text{sim}}$ (PDMS)} \\ \hline 
\multicolumn{ 1}{|c|}{$^1$H} & \multicolumn{ 1}{c|}{$ 0.73$} & \multicolumn{ 1}{c|}{0.77 (\emph{i})} & \multicolumn{ 1}{c|}{0.80 (\emph{iv})} \\ \hline 
\multicolumn{ 1}{|c|}{D} & \multicolumn{ 1}{c|}{$ 0.96^\dagger$} & \multicolumn{ 1}{c|}{0.95 (\emph{ii})} & \multicolumn{ 1}{c|}{0.96 (\emph{v})} \\ \hline
\multicolumn{ 1}{|c|}{$^{19}$F} & \multicolumn{ 1}{c|}{not available} & \multicolumn{ 1}{c|}{0.98 (\emph{iii})} & \multicolumn{ 1}{c|}{0.99 (\emph{vi})} \\ \hline
\end{tabular} 
\caption{Results of experiments ($T_{\text{exp}}$) and simulations ($T_{\text{sim}}$) for nanocomposite model-materials of 100\,$\mu$m thickness containing $^1$H, D or $^{19}$F for the very cold spectrum of Fig.\,\ref{fig2}\,(b). Relative measurement uncertainties are estimated to be up to 5\,\%. ($^\dagger$This value was derived from a measured value stated in Ref.\,\cite{prunerNIMA2006}. See text.)}
\label{tab:Results}
\end{center} 
\end{table}

As expected, measured transmissivity $T_{\text{exp}}$ for thermal neutrons is close to unity (100\%). In fact, due to the rather low precision of the measurement, it cannot be distinguished from unity. The simulation agrees with this result. 

The results of experiments and simulations for VCN are summarized in Table\,\ref{tab:Results}. Fluorinated samples comparable with our holographic nanocomposite gratings used in the experiments were not available. The value $T_{\text{exp}}$ for deuterated PMMA is an estimation based on an experimental value obtained under very different circumstances: It was derived using the Lambert-Beer law \cite{searsBook1989} from the value in Ref.\,\cite{prunerNIMA2006} for three gratings of deuterated pure (not containing nanoparticles) PMMA material of 2.8\,mm thickness each, measured at a wavelength distribution width of 10\,\% centered at 1\,nm. It agrees with the simulation (rightmost column in Table\,\ref{tab:Results}), which indicates that the addition of SiO$_2$ nanoparticles does not play a major role for neutron losses, as expected. 

Most importantly, simulations predict a transmissivity just below unity for $^{19}$F-based nanocomposite grating materials, a very promising result for future application of nanocomposites for neutron optics. Furthermore, the simulation of the angular distribution of a, beforehand, perfectly collimated beam transmitted by a fluorinated PDMS sample predicts that about 98.4\,\% of transmitted neutrons proceed in a circular cone of only $\approx 3.2\times 10^{-3}$ degrees opening angle ($\approx 2.4\times 10^{-9}$\,sr), which is approximately the solid angle of an area of $1\times 1$\,mm$^2$ seen at a distance of 20\,m. To estimate the influence of the rather broad VCN spectrum on transmissivities (for example $T_{\text{sim}}=0.77$, see table), the simulated transmissivity in a monochromatic VCN beam of 37\,{\AA} wavelength was calculated to be $T_{\text{sim}}=0.79$ for conventional PMMA. As expected, the influence is of practical importance insofar as a broad incident wavelength spectrum does not prevent the use of nanocomposite materials.

It has to be considered that neutron scattering cross-sections depend on sample temperature and the wavelength of the incident neutrons (see, for instance, \cite{hillNIMA2005} or \cite{glinkaJAC2011} for the case of hydrogen), which can lead to underestimation of cross sections for low energy neutrons in calculations when using tabulated neutron cross-section data derived from scattering lengths. Such effects are included in our simulations with MCNP.

\section{Conclusion}\label{sec:concl} 

Experimental values confirm our MCNP simulations for transmission of very-cold neutrons through nanocomposite materials that can be applied for diffractive neutron optics applications. It is shown that at our early stage of development the transmissivity of nanocomposite materials containing $^1$H is acceptable for all kinds of proof-of-principle experiments. Furthermore, we find that replacement of $^1$H by deuterium or fluorine atoms will result in a significant decrease in absorption and scattering losses, leading to neutron transmissivity close to 100\,\% even for very-cold neutrons and negligible widening of the beam collimation. This promising result encourages further pursuit of the development of diffractive optical elements for cold and very-cold neutrons based on nanocomposite materials. 
Furthermore, all the results are considered to be valid also for other material classes of similar chemical composition \cite{licenJPCS2017,ellabbanMaterials2017}, which are being tested for use in neutron optics.

\paragraph{{\bf Funding}} 
The authors gratefully acknowledge the financial support of the Centre for International Cooperation \& Mobility of the Austrian Agency for International Cooperation in Education and Research (grant no. SI 15/2018 or BI-AT/18-19-009), and the hospitality of the Institut Laue Langevin. 

\section*{References}

\bibliography{klepp.bib}
\end{document}